\documentclass{elsart}
\usepackage{graphics}
\usepackage{graphicx}
\newcommand{\ve}[1]{\mathbf{#1}}

\newcommand{\be}{\begin{equation}}
\newcommand{\ee}{\end{equation}}
\newcommand{\bea}{\begin{eqnarray}}
\newcommand{\eea}{\end{eqnarray}}

\newcommand{\bc}{\begin{center}}
\newcommand{\ec}{\end{center}}

\newcommand{\beq}{\begin{eqnarray*}}
\newcommand{\eeq}{\end{eqnarray*}}

\def\hsp5{\hspace{5mm}}

\def\case#1/#2{\textstyle\frac{#1}{#2}}

\usepackage{graphicx}
\usepackage{dcolumn}
\usepackage{bm}

\begin{document}

\begin{frontmatter}

\title{An analysis of cross-correlations in an emerging market}

\author[label1,label3]{Diane Wilcox}
\ead{diane@maths.uct.ac.za}
\author[label2]{Tim Gebbie}
\ead{tim.gebbie@physics.org}

\thanks[label3]{Corresponding author.}

\address[label1]{Dept. of Mathematics \& Applied Mathematics, University of Cape Town, Rondebosch, 7700, South
Africa}
\address[label2]{Futuregrowth Asset Management, Private Bag X6, Newlands, 7725, South Africa}

\begin{abstract}
We apply random matrix theory to compare correlation matrix
estimators $C$ obtained from emerging market data. The correlation
matrices are  constructed from 10 years
  of daily data for stocks listed on the Johannesburg Stock Exchange
  (JSE) from January 1993 to December 2002. We test the
 spectral properties of $C$ against random matrix predictions and find
 some agreement between the distributions of eigenvalues, nearest
 neighbour spacings, distributions of eigenvector components and the inverse participation ratios for
 eigenvectors.   We show that interpolating both missing data and
 illiquid trading days with a zero-order hold increases agreement with RMT predictions.
 For the more realistic estimation of correlations in an emerging market, we suggest
 a pairwise measured-data correlation matrix. For the data set used, this
 approach suggests greater temporal stability for the leading
 eigenvectors. An interpretation of eigenvectors in terms of trading strategies
is given, as opposed to classification by economic sectors.
\end{abstract}


\end{frontmatter}

\section{Introduction}

Correlation matrices are common to problems involving complex
interactions and the extraction of information from series of
measured data. Our aim is to determine empirical correlations in
price fluctuations of  daily sampled price data of distinct shares
in a reliable way. Our investigation is based on 10 years of daily
data for 250-350 traded shares listed on the JSE Main Board from
January 1993 to Dec 2002.

There are several aspects to the question of how to calculate
correlations in financial time series. In particular, missing data
and thin trading (no prices changes for a stock over several time
periods) may be significant. Random correlations in price changes
are likely to arise in an ensemble of several shares. Furthermore,
for a portfolio of $N$ distinct assets, there will be $N(N-1)/2$
entries in a correlation matrix which has been determined from
time series of length $L$. When $L$ is not large, the calculated
covariance matrix may be dominated by measurement noise. Hence, it
is necessary to understand effects of (i) noise (ii) finiteness of
time series (iii) missing data and (iv) thin trading in
determination of empirical correlation.

The properties of random matrices first became known with Wigner's
seminal work in the 1950's for application in nuclear physics in
the study of statistical behaviour of neutron resonances and other
complex systems of interactions (\cite{mehta}, \cite{brody} and
\cite{guhr}). More recently  random matrix theory  has been
applied to calibrate and reduce the effects of noise in financial
time series and  to investigate constraints on rational
(empirically based) decision making (cf. \cite{physa259},
\cite{prl83a},  \cite{pre63a}, \cite{pre65a}, \cite{DrGrRuSp}
\cite{physa343}, \cite{drKwSpWo1}, \cite{DrKwGrRuSp1} ).
Correlation matrices are computed for the data under investigation
and quantities associated with these matrices may be compared to
those of random matrices. The extent to which properties of the
correlation matrices deviate from random matrix predications
clarifies the status of the information derived from the
computation of covariances. In several studies of shares traded in
the S\&P 500 and DAX, it was found that, aside from a small number
of leading eigenvalues, the eigenvalue spectra for the measured
data coincide with theoretic random matrix predictions, i.e. it
was found that the estimation of covariances is dominated by
random noise.  In \cite{noh}, postulated a model for the
correlations which explained the observed spectral properties. RMT
has also been shown to yield an improved estimation technique: an
estimated correlation matrix can be filtered by removing the
contributions of eigenvalues which lie in the RMT range. In
\cite{physa319} it is shown that noise levels in the correlation
matrix depend on the ratio $N:L$, where $N$ denotes the number of
stocks and $L$ denotes the length of the time-series.

\subsection{Correlation matrices and missing data in an emerging
market}

In this paper we consider the problems of missing data and thin
trading in determination of empirical correlation in daily sampled
price fluctuations. We analyze the data base containing prices
$S_i(t)$, the prices of assets $ i = 1, \ldots , N$  at time $t$
as follows. We first find the change in asset prices \beq r_i(t) =
\ln\,S_i(t+\triangle t ) - \ln\,S_i(t). \eeq

The {\em usual cross-correlation matrix} for idealized data
(non-zero price fluctuations and no missing data) is given by \beq
C_{ij} := \frac{\langle r_ir_j\rangle - \langle r_i\rangle\langle
r_i\rangle}{\sigma_i\sigma_j},
 \eeq

 where$ \langle \ldots \rangle $ denotes average over period
 studied and $ \sigma_i ^2:= \langle r_i^2\rangle -
 \langle r_i\rangle ^2 $ is the variance of the price changes of
 asset $i$.
Alternatively one could write \beq C_{ij} =
\frac{1}{L}\sum\limits_{t=1}^{L}{R_i(t)R_j(t)} \eeq where $L$
denotes the uniform length of the time series and  $R_i(t) $
denotes the price change of asset $i$ at time $t$ such that the
average values of the $R_i's$ have been subtracted off and the
$R_i's$ are rescaled so that they all have constant volatility $
\sigma_i ^2:= \langle R_i^2\rangle = 1. $ This is written as
$\mathbf{C} = \frac{1}{L}\mathbf{M}\mathbf{M}^T$ where $
\mathbf{M}$ is a $N\times L$ matrix and $\mathbf{M}^T$ is its
transpose (cf. \cite{prl83a}).

The {\em pairwise measured-data cross-correlation matrix} using
the {\em pairwise deletion method} \cite{Wo}, \cite{missing}) for
the case when there is missing data in time series of returns is
computed as follows: \beq {\mathcal C}_{ij} := \frac{\langle
\rho_i \rho_j\rangle - \langle \rho_i\rangle\langle
\rho_i\rangle}{\sigma_i\sigma_j},
 \eeq

where $\rho_i$ and $\rho_j $ denote subseries of $r_i$ and $r_j$
such that there  exists measured data for both $\rho_i$ and
$\rho_j $ at every time period in the subseries, and $ \langle
\ldots \rangle $ denotes average over period
 studied, $ \sigma_i ^2:= \langle \rho_i^2\rangle -
 \langle \rho_i\rangle ^2 $ is the variance of the price changes of
 asset $i$.

\section{Random Matrix Theory (RMT) predictions}

We summarise four known universal properties of random matrices,
namely the Wishart distribution for eigenvalues, the Wigner
surmise for eigenvalue spacing, the distribution of eigenvector
components and the inverse participation ratio for eigenvector
components, which will be applied in our analysis.


Let $A$ denote an $N\times L$ matrix whose entries are i.i.d
random variables which are normally distributed with zero mean and
unit variance. As $N,L \rightarrow \infty$ and while $Q= L/N$ is
kept fixed, the probability density function for the eigenvalues
of the Wishart matrix  (or Laguerre ensemble) $R =
\frac{1}{L}AA^T$ is given by (\cite{bai}, \cite{edelman},
\cite{pre60a}):

\bea p(\lambda) & = & \frac{Q}{2\pi} \frac{\sqrt{(\lambda_{\max}
-\lambda) (\lambda-\lambda_{\min})}}{\lambda} \label{wishart} \eea
for $\lambda$ such that $ \lambda_{\min}  \leq   \lambda \leq
\lambda_{\max}$, where $ \lambda_{\min}$ and $ \lambda_{\max}$
satisfy

\bea
 \lambda_{\max / \min} = & 1+\frac{1}{Q} \pm 2\sqrt{1/Q}. \label{minmax} \eea

The distribution of eigenvalue spacings was introduced as a
further test for the case when the empirical eigenvalue
distribution does not deviate significantly from the RMT
predication. The so-called {\em Wigner surmise} for eigenvalue
spacings \cite{mehta} is given by \bea p(s) =
\frac{s}{2\pi}\,exp\,(- \frac{s \pi^2}{4}), \label{surmise} \eea
where $s = (\lambda_{i+1} - \lambda_{i})/d$ and $d$ denotes the
average of the differences $\lambda_{i+1} - \lambda_{i}$ as $i$
varies \footnote{Unfolded eigenvalues are used in practice
\cite{brody}, \cite{guhr} \cite{pre65a}. }.


It has been found that the eigenvector components $v_a^i$ for $ a
= 1 \ldots n $ of an eigenvector  $\ve{v}_a$  are normally
distributed with  zero mean and unit variance \cite{guhr},
\cite{boupot},

\bea p(u) = \frac{1}{\sqrt{2 \pi}}\,exp\,(-\frac{u^2}{2}).
\label{porthom} \eea

The inverse participation ratio (IPR) is used to analyze the
structure of the eigenvectors of the correlation matrix
\cite{pre65a}. The $i$th component ${v_a^i}$  of  $\ve{v}_a$
corresponds to the contribution of the $i$th time series to that
eigenvector. To quantify this contribution, the IPR for $\ve{v}_a$
is defined \bea I_a = \sum_{i=1}^N (v_a^i)^4, \label{ipr} \eea
where $N$ is the number of time series (the number of shares) and,
hence, number of eigenvalue components. If the components of the
eigenvector are identical, $v_a^i = \frac{1}{\sqrt{N}}$, then $I_a
= 1/N$; if there is only a single non-zero component, then $I_a =
1$. In general, the IPR is a reciprocal of the number of
eigenvector components which are contribute significantly, i.e.
which are different from zero. It is found that $E[I_a] = 3/N$
since the kurtosis for the distribution of eigenvector components
is $3$.

\section{Analysis of Johannesburg Stock Exchange data}

The JSE is one of the  20 largest national stock markets in the
world. We summarise some of its known qualitative features.
Although many of the main board JSE shares are illiquid, the
market as a whole is a fairly liquid one. There is share
concentration in half-dozen shares: these dominant shares account
for almost a third of the index and have a large bias towards
resources. The resources sector in turn is strongly correlated
with the dollar-rand exchange rate, an exogenous factor that has a
dominant influence on price dynamics in South African stock
markets.  Next it is noteworthy that different shares are listed
on the JSE at different times and, hence,  different shares do not
always trade on the same day. However,  some shares which do not
trade often may occasionally  trade in large volumes for several
days. These realities exacerbate the problem of estimating
correlations in a reliable way.

The data set used in this study incorporated  a zero-order hold
for prices when there was no trading. This approach accounts for
sequences of zero-valued returns in the return times-series even
though no measurements occurred. While it has often been
convenient to set the returns to zero in the periods preceding
listing of shares to avoid data holes, in general, this strategy
seems to give rise to a significant gaussian component to
estimated correlations. We investigate the effect of various
treatments and interpretations of measurements in the context of
price time-series. The approach here favours the notions that (1)
if no price was discovered for a given share then there was no
measurement, and (2)  share cross-correlations can only be
computed when there are measurements on the same day.

\subsection{Filtering and partitioning the data}

The data set of 10 years of data from 1 January 1993 to 31
December 2002 was split into annual epochs. The data was windowed
to create 6 sets of 5 years of daily price data. Each block was
screened to remove shares that were de-listed or which traded
quite infrequently. For each year in a given epoch of 5 years,
this was achieved by dropping all shares that neither recorded
price measurements at year-end nor traded at least once in the
preceding month. Table \ref{tab:dataset} gives the data sets used.


\begin{figure}
  \centering
  \includegraphics[width=16cm]{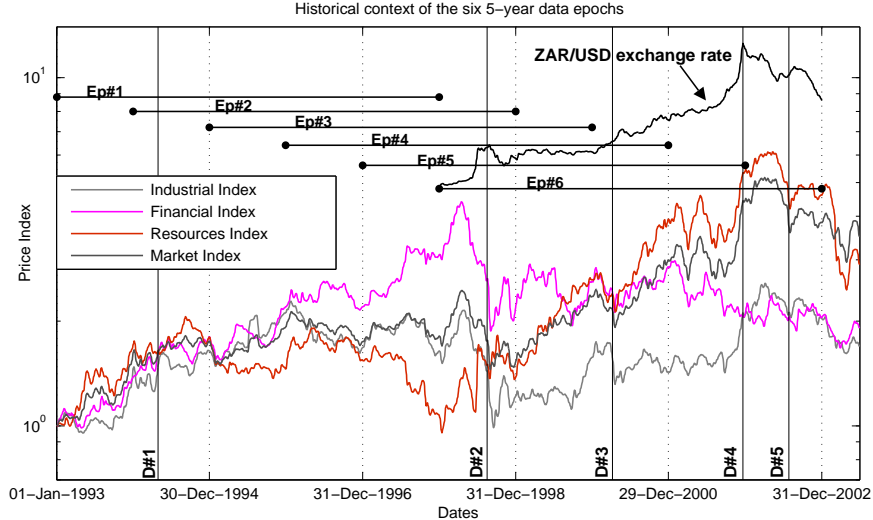}
  \caption{Historical context of price evolution for the period
  studied: Indices are reconstructed using the market
  capitalization of stocks in the industrial, financial and
  resources sectors as well as the entire market for the market
  index. The horizontal lines labelled EP$\#$1, EP$\#$2 to EP$\#$6
  demarcate time windows 1993-1997, 1994-1998 to 1998-2002,
  respectively.
  Specific dates are highlighted by vertical lines labelled:
  D$\#$1 - 27 Apr 1994 - the first SA elections of the post-Apartheid era,
  D$\#$2 - 17 Aug 1997 - Russian GKO default,
  D$\#$3 - 10 Apr 2000 - proxy date for Nasdaq crash,
  D$\#$4 - 20 Dec 2001 - SA Rand (ZAR) crash
  D$\#$5 - 27 Jul 2002 - Sarbanes-Oxley Act.
  Inset: The ZAR/USD exchange rate evolution from
  Jan 1998 to Dec 2002. }\label{fig:hist}
\end{figure}

In Figure ~\ref{fig:hist} we reconstruct price indices
  (not total price) from the market capitalization of each
  individual stock based on the economic sector membership, i.e. a
  weight in a particular index would be the stocks market capitalization
  divided by the total portfolio market capitalization. The reason for this
  is that there is no complete constituent history available over
  for the full 10 year period studied
  - the indices were reconstructed by the authors. This also
  ensures consistency between the indices provided and the stocks
  and the stock data used in the study. Figure ~\ref{fig:hist}
 corroborates evidence of negative correlation which is
  presented  when we consider temporal stability of our results in
  Subsection ~\ref{tempstab} below - the dominance of the financial sector
  peaks in 1998 for the period under investigation; thereafter
  the resources sector begins to dominate the market.

\subsection{Three estimates of cross-correlations}

 We investigate correlation structure by considering correlation
matrices of the data sets in Table \ref{tab:dataset} in three
different ways. Case 1:  we assign the value of zero whenever
there is no measured data for a return $r_i(t)$ for asset $i$ at
time $t.$; we then compute the correlation matrix in the usual way
as described in the introduction. Case 2:  we compute the {\it
pairwise measured-data correlation } matrix to overcome missing
measurements. Case 3: we address the problem of no trading, i.e.
zero price fluctuations for several time periods in succession. To
do so, in the event of $2$ or more successive zero-valued price
fluctuations we delete the measured return value $r_i(t)=0$,
effectively turning the zero-valued information into missing data.
This compensates for interpolated prices being mistaken for
measurements. We then compute the pairwise measured-data
correlation matrix.

We considered the problem of non-positive definiteness (this
property is destroyed by most missing-data methods, including the
one which we implement) by applying the area-minimizing algorithm
of \cite{ChMc}  to make the matrices in Case 3 positive
semi-definite (see also \cite{Lindskog}). Lastly, as a further
case, we removed bias from the data as a means of removing the
market mode (cf. \cite{pre65a}). Details for these cases are not
included since they do not add to our discussion on the
phenomenology on missing data in this paper.

We note that there are several other methods for treating missing
data (see for example \cite{Wo} and \cite{missing}). Our view is
that the {\em pairwise deletion} method offers a sufficiently
robust correlation estimate for the purposes of this analysis of
daily data. {\em Listwise deletion} of an entire day's records for
a day on which there is missing data for a single stock is likely
to delete useful data for remaining stocks and results in too few
remaining records. {\em Mean substitution} and {imputation by
regression} are likely to introduce spurious correlations between
stocks which are not listed or which do not trade for several
successive periods - this is borne out by the comparison of
results between Cases 2 and 3. For the application of {\em hot
deck imputation} it is not at all clear which strings of return
data it would be appropriate to draw from over the varying time
windows. The most promising alternative missing data method would
be the use of an {\em expectation maximization algorithm} and this
is left for further investigation.

\begin{figure}
  \centering
  \includegraphics[width=10cm]{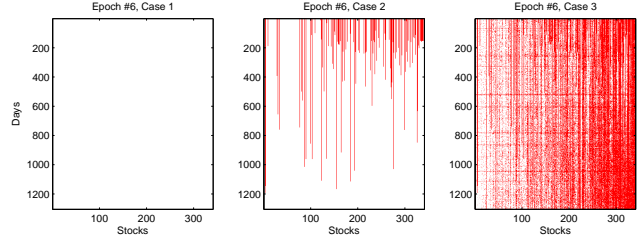}
  \caption{Missing data is depicted in red, while reasonable data is white space.
  The three graphs are, from left to right are: Case 1, which includes zero-padding and zero-order hold,
  Case 2 which has only zero-order hold, and Case 3 which has measured data only.
  These demonstrate the extent of missing data. The graph reflects daily
  data from Epoch 6 (1998-2002).}\label{fig:missdata}
\end{figure}

The different treatments of the data have significant impact on
the relationship between the number of meaningful data points and
the normalisation factors used in the computations.

Figure ~\ref{fig:missdata} illustrates the occurrence of missing
data. In Case 1, zero padding fills all the gaps so there appears
to to be no missing data. In Case 2, there is missing data for
shares which were not listed at the start of the epoch and
constant prices are recorded when no trading has taken place. In
case 3, we remove prices when no trading has occurred. Stocks are
ordered by market capitalization from left to right. The
concentration of red on the right of the (c) is clear evidence
that smallest cap stocks tend to trade quite infrequently.
Horizontal bands indicate public holidays. That zero order hold
was used to interpolate missing data on public holidays within the
data set obtained, is a prime example of how measurement error can
contaminate a database.

\begin{figure}
  \centering
  \includegraphics[width=10cm]{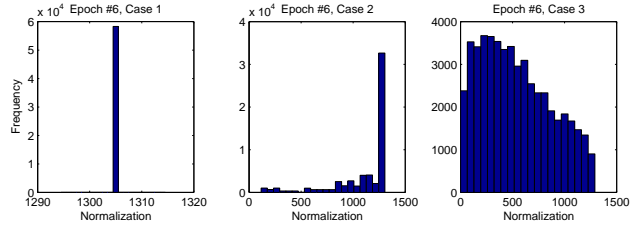}
  \caption{Frequency distributions of normalisations used in the
  computation of pairwise correlations
  are plotted as a histogram for each Case considered in Figure
  ~\ref{fig:missdata}. See also Table ~\ref{tab:dataset}.
}\label{fig:normfreq}
\end{figure}

Figure ~\ref{fig:normfreq} plots the frequency of the
normalisation factors used to compute entries in the correlation
matrices. For cases 2 and 3, incorrect normalisation factors for
(truncated) pairwise matched timeseries would have the effect of
distorting correlation estimates when there is missing data. In
Case 1, the normalisation factor is the same for all pairs, i.e.
it is equal to $L\approx 1305$ , the total number of official
trading days in the 5-yr epoch (see table). In Case 2, the
normalisation factor is often much larger than the number of days
for which shares are actually traded. Here, normalisations in the
order of $L$ were frequently used even though there was
substantially less measured price data. In Case 3, normalisation
factors varied from less than 60 up to $L$, with normalisations of
330-390 occurring most frequently.

{\small
\begin{table}
\caption{\label{tab:dataset}The data sets used comprising shares
traded on the Johannesburg Stock exchange. Each data set starts on
the 1 January of the starting year and ends on the 31 December of
the ending year.}
\begin{tabular}{llcccccc}
\hline
    & Data set (Epoch) & 1 & 2 & 3 & 4 &  5 & 6\\
\hline
    & Start date (1 Jan)  & 1993 & 1994 & 1995 & 1996 & 1997 & 1998\\
    & End date   (31 Dec) & 1997 & 1998 & 1999 & 2000 & 2001 & 2002\\
    & Total no. of shares (N) & 253  & 296  & 321  & 330  & 336  & 341\\
    & No. of trading days (L)    & 1304 & 1305 & 1306 & 1306 & 1305 & 1305\\
\hline
          & No. of shares used & 253  & 296  & 321  & 330  & 336  & 341  \\
Case 1    & \% zero returns    & 73\% & 70\% & 63\% & 58\% & 53\% & 54\% \\
          & \% missing data    & 0    & 0    & 0    & 0    & 0    & 0   \\
\hline
          & No. of shares used & 253  & 296  & 321  & 330  & 336  & 341  \\
Case 2    & \% zero returns    & 54\% & 44\% & 41\% & 40\% & 42\% & 46\% \\
          & \% missing data    & 18\% & 24\% & 23\% & 18\% & 13\% &  8\% \\
\hline
          & No. of shares used & 244  & 282  & 308  & 310  & 316  & 319  \\
Case 3  & \% zero returns    & 12\% & 11\% & 12\% & 14\% & 14\% & 15\% \\
          & \% missing data    & 59\% & 56\% & 50\% & 42\% & 38\% & 36\% \\
\hline
\end{tabular}
\end{table}
}

\vspace{0.5cm}

\begin{figure}
  \centering
  \includegraphics[width=12cm]{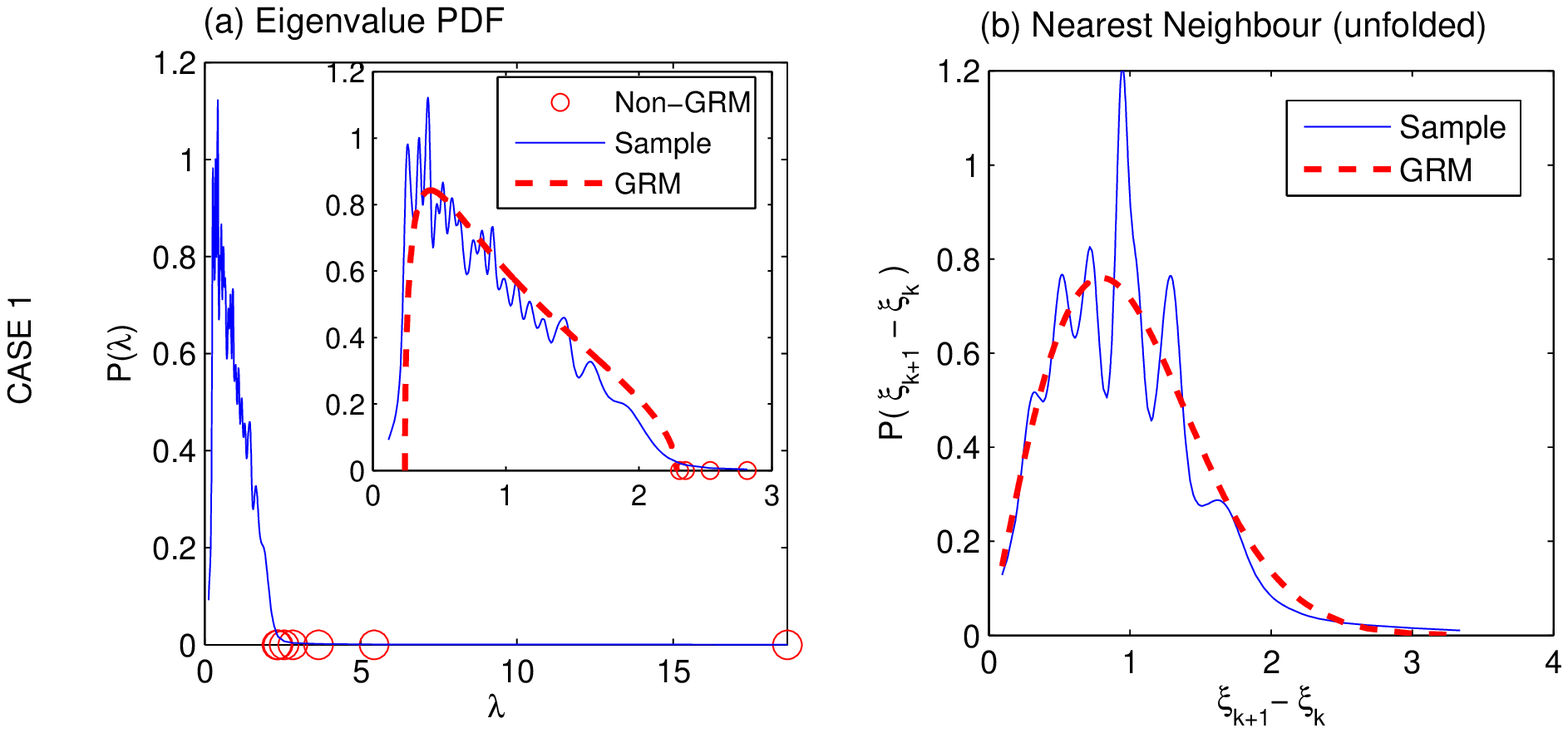}
  \includegraphics[width=12cm]{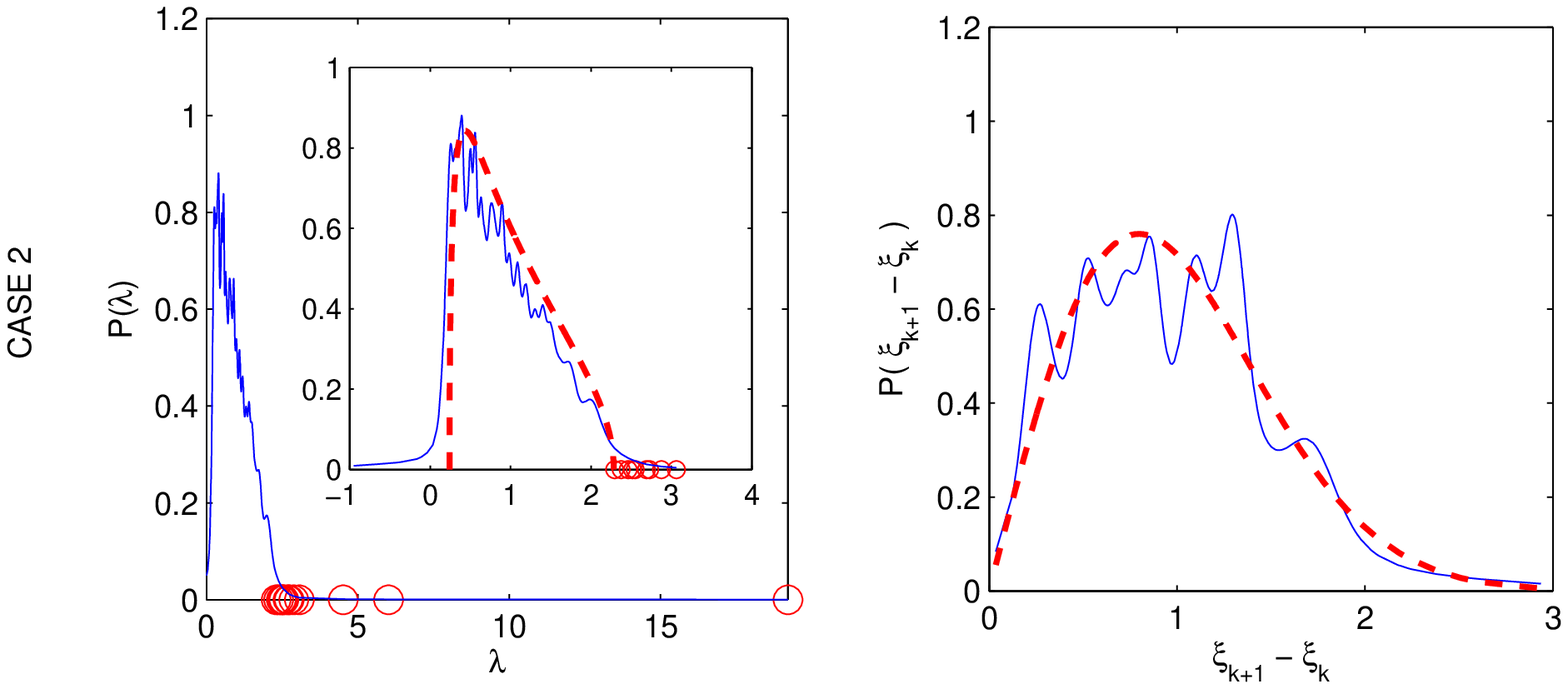}
  \includegraphics[width=12cm]{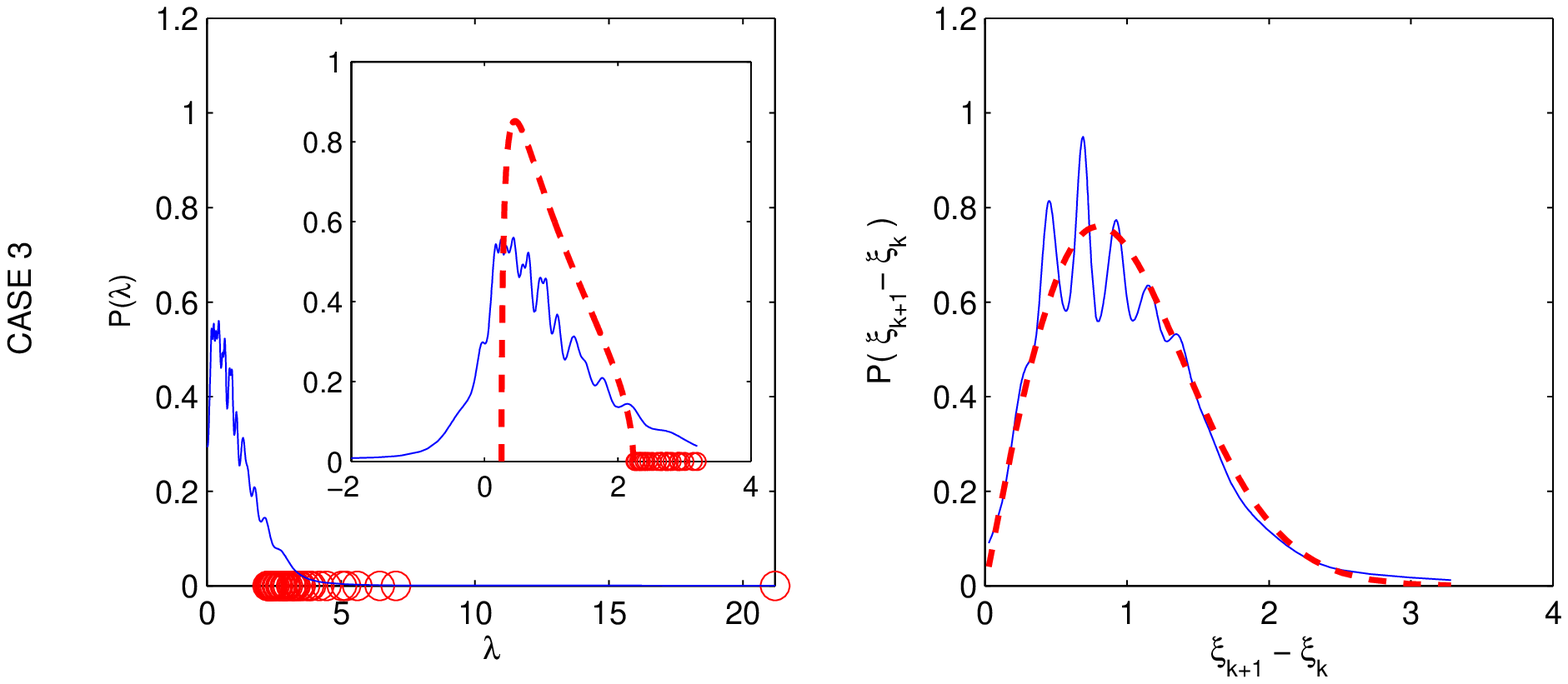}
    \caption{Daily price returns for JSE main board shares for years 1998-2002 are used to investigate
  eigenvalue structures of  three estimated correlation matrices.  Figure ~\ref{fig:values} (a)  shows the eigenvalue density functions with
  the distinct eigenvalues greater than the maximum RMT predicted value for the same Q-factor as the sample.
  Insets: plots of the Wishart distribution (Eqn.~\ref{wishart}) are superimposed on plots of the small eigenvalues.
 Figure ~\ref{fig:values} (b) shows the nearest-neighbour distributions of
  the folded eigenvalues. Superimposed on these are plots of the Wigner Surmise
 ( Eqn.~\ref{surmise}). The folded eigenvalues were computed using Gaussian
  broadening and numerical integration.
  }\label{fig:values}
\end{figure}

\subsection{Spectral properties and comparisons with RMT predictions}

In \cite{prl83a}, investigating daily price fluctuations for N=406
stocks of the S\&P 500 for L=1309 days during 1991-1996, with Q =
3.22, it was found that the leading eigenvalue was $\approx$25
times greater than the RMT predicted $\lambda_{max}$. Adjusting
for the total variance $\sigma_2$ of the price fluctuations, it
was found that 94\% of the spectrum could be attributed to random
noise. In \cite{prl83b}, the high-frequency TAQ database published
by NYSE for the period 1994-1995 was analysed: using 30 min
returns for N=1000 companies with L=6448, it was found again that
the leading eigenvalue was $\approx$25 times larger than the RMT
predicted value for $\lambda_{max}= 1.94$ and that $\approx$98\%
of the eigenvalues could be accounted for as random noise effects.
Results of \cite{prl83b} were corroborated in the more extensive
study \cite{pre65a} of the same high-frequency TAQ data together
with CRSP databases of daily data for common stocks in the NYSE
beginning 1925, the AMEX beginning 1962 and the NASDAQ beginning
1972. Investigation of further universal properties confirmed that
for eigenvalues within the Wishart range and their corresponding
eigenvectors:   (a) the distribution of nearest-neighbour spacings
were in good agreement with Gaussian Orthogonal Ensemble
predictions, (b)
 the distribution of eigenvector components conformed with the
predicted Gaussian distribution with zero mean and unit variance
and (c)  almost all the eigenvector components contributed equally
to the inverse participation ratio except for eigenvectors
corresponding to eigenvalues outside the RMT bounds. In the latter
cases it was found that almost all stocks participated in the
largest eigenvector while for the remaining large eigenvectors
there was localization, i.e. only a few stocks contributed to
them. Similar analysis was conducted on high-frequency data for
the DAX for the period Nov 1997 to Dec 1998 to examine intraday
dynamics and memory effects in the index \cite{DrKwGrRuSp1}.

In this section we investigate the same properties in the context
of an emerging market. Figure ~\ref{fig:values}  gives a
comparison of the eigenvalue densities and nearest-neighbour
spacings for the three different cases considered for the last
epoch, 1998-2002. It is clear from the graph that most of the
eigenvalues for Case 1 are within the range of the Wishart
distribution (Eqn.~\ref{wishart}). For Case 2, the number of
eigenvalues within the noise range is slightly reduced. Some of
the eigenvalues are negative in this case. For Case 3 there is a
more significant drop in number of eigenvalues in the noise band
compared to Case 1; there are also more negative eigenvalues in
Case 3 compared to Case 2. In all Cases the nearest-neighbour
spacings indicate some agreement with the Wigner surmise
(Eqn.~\ref{surmise}).

 It is clear from the presence of
negative eigenvalues in Cases 2 and 3 that the matrices obtained
are no longer positive definite. There are several algorithms to
obtain positive definite matrices from a non-positive definite
 estimate ~\cite{ChMc}, \cite{Lindskog}. A thorough
comparison of these methods (including their impact on eigenvalue
distributions and temporal stability) is a separate topic of
investigation.

For Case 3 we found that $\approx$88\% of the eigenvalues
(including negative values) fell below $\lambda_{max}= 2.23$ (a
smaller percentage than in \cite{prl83a}, \cite{prl83b} and
\cite{pre65a}) and that the largest eigenvalue, $\lambda = 21.20$,
was $\approx$9.5 times great than $\lambda_{max}$ (significantly
less than results for developed markets \cite{prl83a},
\cite{prl83b} and \cite{pre65a}). The high percentage of
eigenvalues below $\lambda_{min}$ may be attributed to the fact
that many of the less liquid stocks behave independently relative
to the rest of the market. While a null-hypothesis of Gaussian
returns is useful for identifying how zero-padding and zero-order
hold add noise to data, it  is possible that the noise range
$[\lambda_{max}, \lambda_{min}]$ is wider than suggested by
equation (~\ref{wishart}). Simulations for time-series with
Gaussian returns populated appropriately with missing data yield
almost identical distributions of eigenvalues as the Wishart
distribution. Different stocks in the SA market exhibit a range of
return phenomenologies, including periodic and aperiodic behaviour
\cite{WiGe}; hence, the construction of a representative
null-hypothesis becomes problematic. From qualitative information
about the market, results for inverse participation ratios,
temporal stability and style characteristics (discussed in
sections below) and by an analysis of the dimensionality of the SA
market \cite{PoGe}, a more realistic estimate seems to be that 8-9
eigenvalues are associated with information content ($\approx$2\%
of the total).

 {\small
\begin{table}
\caption{\label{tab:variances} Average percentage of variance
explained by the leading eigenvalues (average taken over 6
epochs).}
\begin{tabular}{lccc}
\hline
    & Case 1 & Case 2  & Case 3 \\
\hline
Total variance                   & & \\
\% explained by eigenvalues $1-5$  & \ $27\,\% \ \pm\,5\,\%$ & \ \ $25\,\% \ \pm\,4\,\%$ & \ \ $28\,\% \ \pm\,4\,\%$ \\
\% explained by eigenvalues $1-15$ & \ $43\,\% \ \pm\,5\,\%$ & \ \ $41\,\% \ \pm\,4\,\%$ & \ \ $50\,\% \ \pm\,4\,\%$ \\
\hline
Trace of Correlation matrix       & & \\
\% explained by eigenvalues $1-5$  & \ $11\,\% \ \pm\,1\,\%$ & \ \ $12\,\% \ \pm\,1\,\%$ & \ \ $15\,\% \ \pm\,1\,\%$ \\
\% explained by eigenvalues $1-15$ & \ $18\,\% \ \pm\,2\,\%$ & \ \ $20\,\% \ \pm\,2\,\%$ & \ \ $25\,\% \ \pm\,2\,\%$ \\
\hline
\end{tabular}
\end{table}
}

\vspace{0.5cm}

\begin{figure}
  \centering
  \includegraphics[width=12cm]{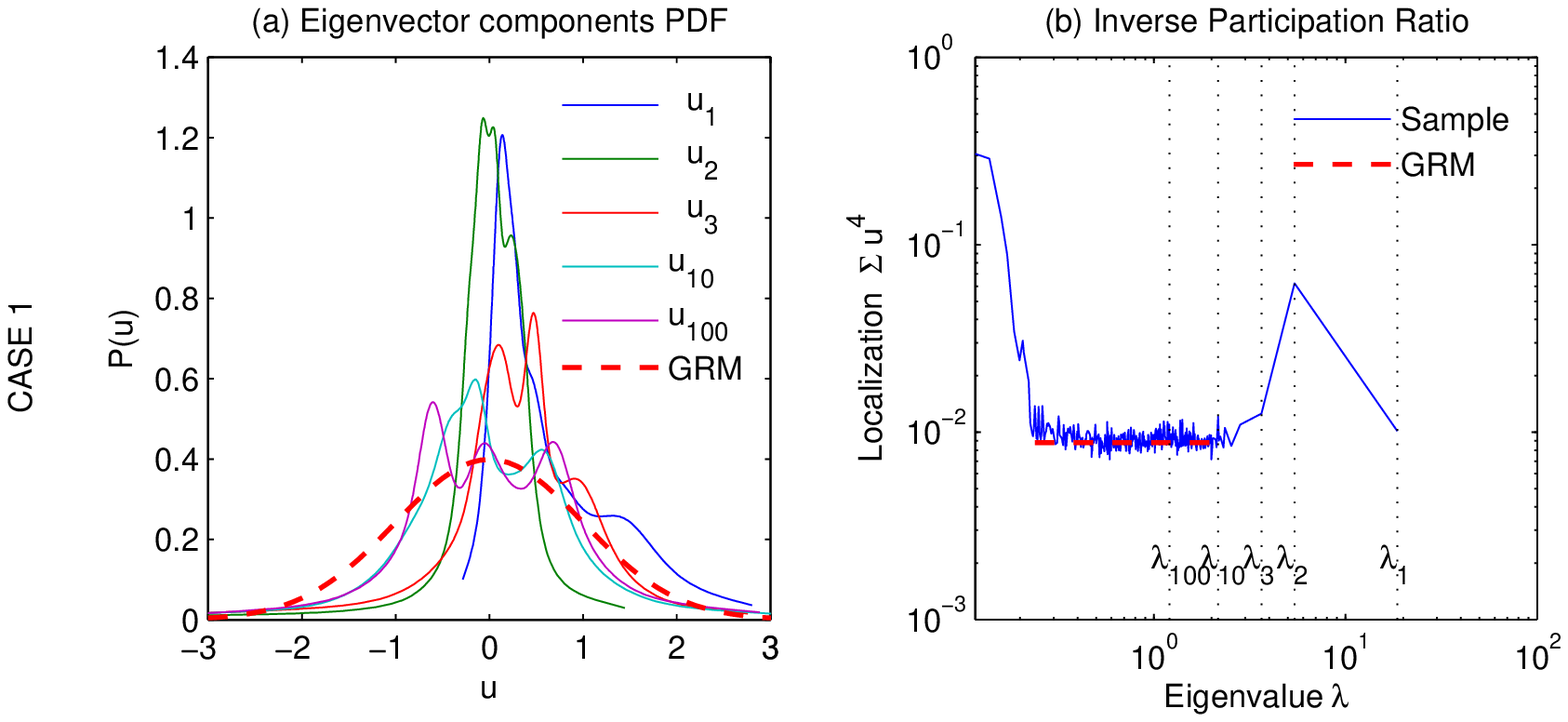}
  \includegraphics[width=12cm]{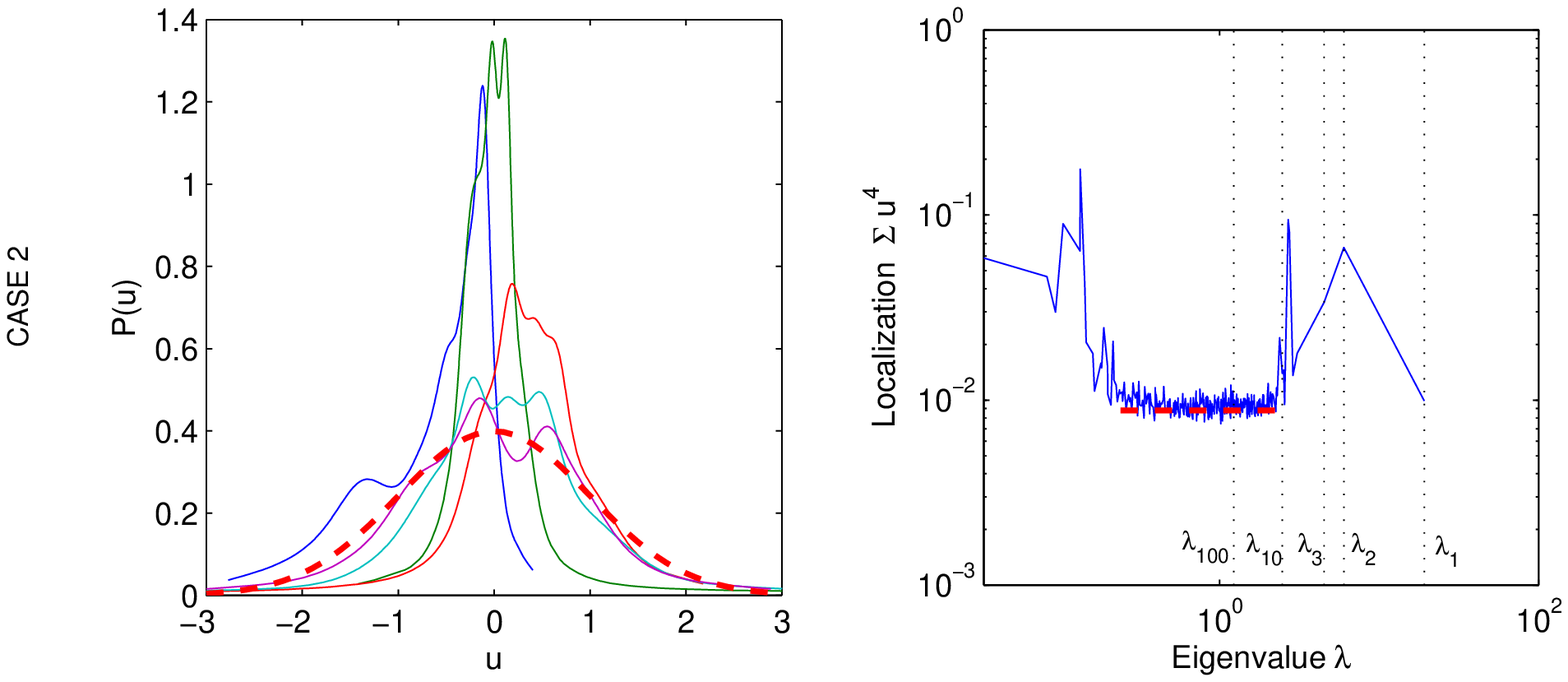}
  \includegraphics[width=12cm]{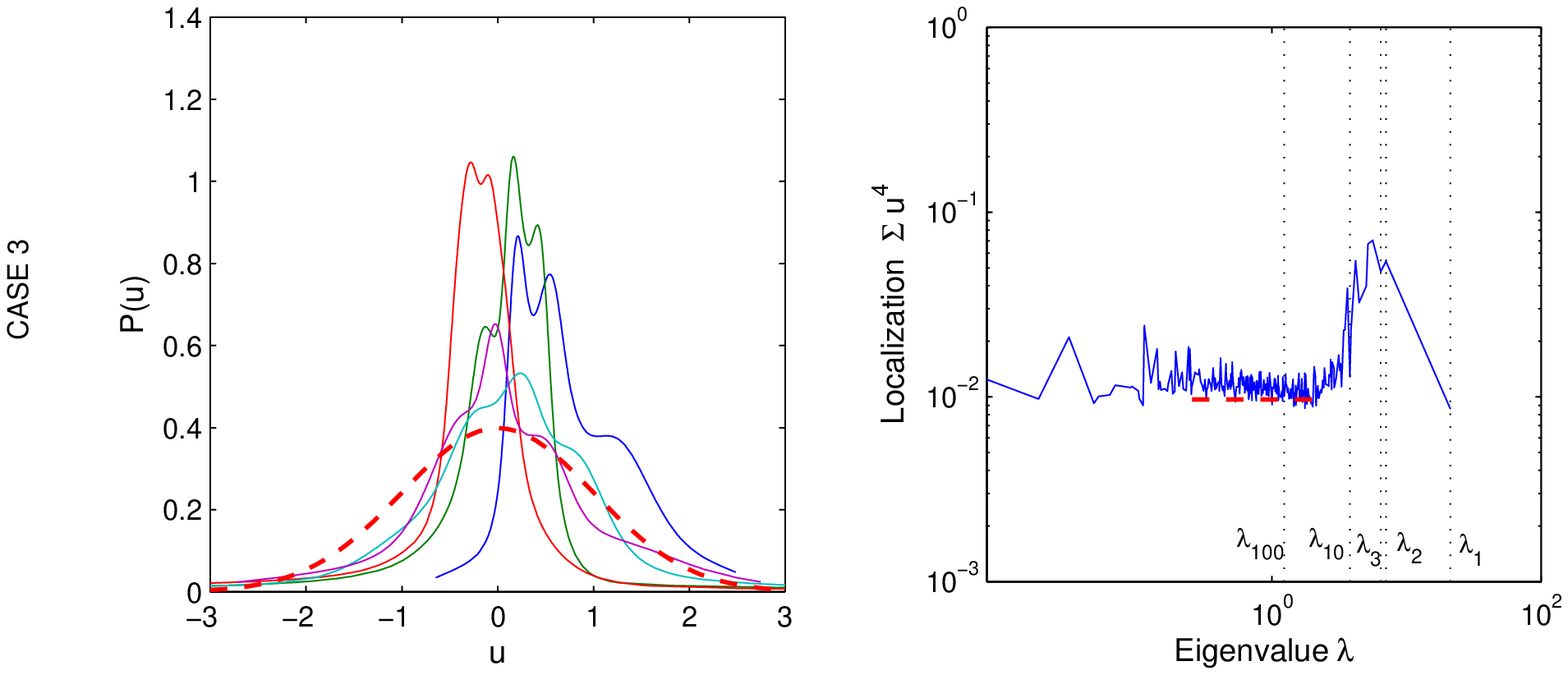}
  \caption{Daily price returns for JSE main board shares for years 1998-2002 are used to investigate
  eigenvectors of  three estimated correlation matrices.
  Figure ~\ref{fig:vectors} (a)
 gives the distributions of
component values for eigenvectors corresponding to the $1^{\it
st}$, $2^{\it nd}$, $3^{\it rd}$, $10^{\it th}$ and $100^{\it th}$
largest eigenvalues. Plots of the Porter-Thomas distribution
  (Eqn. \ref{porthom}) are superimposed.
  Figure ~\ref{fig:vectors} (b) gives plots of the inverse participation ratios
  for the three cases together with the RMT prediction, $E[I_a] =
  3/N$, using kurtosis of Eqn.\ref{porthom}. }\label{fig:vectors}
\end{figure}

Figure ~\ref{fig:vectors} (a)  gives the distributions of
component values for eigenvectors corresponding to the $1^{\it
st}$, $2^{\it nd}$, $3^{\it rd}$, $10^{\it th}$ and $100^{\it th}$
largest eigenvalues for the last epoch, 1998-2002. In all three
cases the first three distributions deviate significantly from the
Porter-Thomas null-hypothesis (Eqn.~\ref{porthom}); the
distributions corresponding to smaller eigenvalues are in greater
agreement with their random matrix counterparts. In Case 3, the
components for the leading eigenvector are mostly positive valued
(as in \cite{prl83a}, \cite{prl83b} and \cite{pre65a}).

Figure ~\ref{fig:vectors} (b) gives the inverse participation
ratios (IPR's) plotted against corresponding eigenvalues. In all
three cases, the IPR for the leading eigenvector is approximately
equal to the RMT prediction of 3/N (Eqn.~\ref{ipr}), indicating
contributions from almost all stocks  in the market. This is
consistent with findings in \cite{pre65a}. For Case 1, the IPR's
for the $2^{\it nd}$ and $3^{\it rd}$ largest and the 7 smallest
eigenvalues deviate significantly from the random matrix null
case. For Cases 2 and 3, the IPR's for largest 6 and 9
eigenvalues, respectively, deviate significantly from the random
matrix null case, indicating contributions from only a few stocks
(as in \cite{pre65a}); the same is true for several of the
smallest eigenvalues for Case 2; for Cases 2 and 3, the rest of
the eigenvalues, the mean IPR is greater than the RMT prediction
of 3/N  and IPR values fluctuate with greater variance about this
mean compared with variance for the null case.

\begin{figure}
  \centering
  \includegraphics[width=8.5cm]{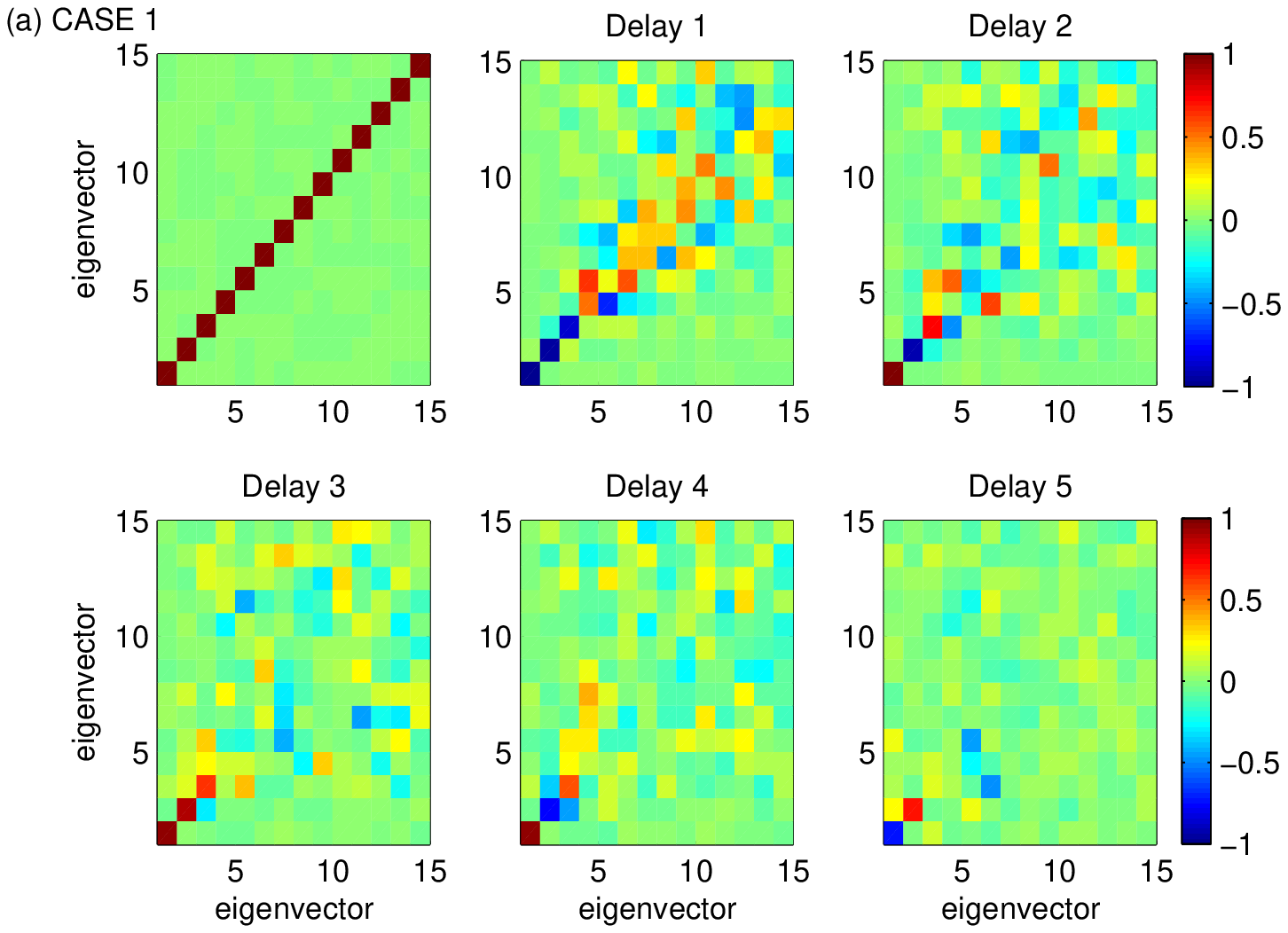}
  \includegraphics[width=8.5cm]{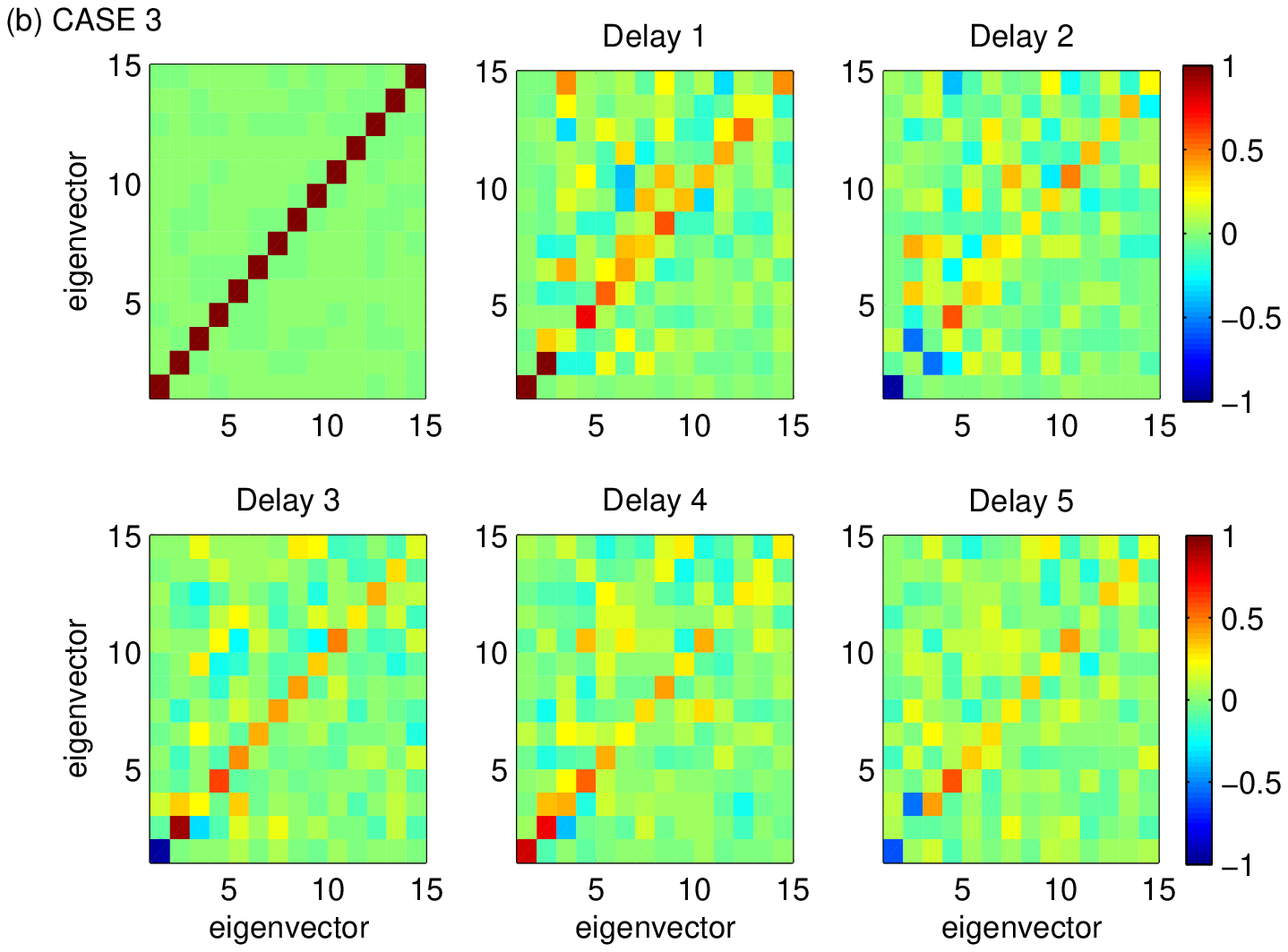}
  \caption{Overlap matrices are computed using 6 sets of 5-yr epochs of daily returns.
  Figure ~\ref{fig:overlap1} (a) and (b) give results for Cases 1 and
  3, respectively. The graphs depict estimated correlations between the 15 leading eigenvectors from
the last epoch, $1998-2002$, with their analogues from the
preceding epochs, $1998-2002$ (lag 0),  $1997-2001$ (lag 1) to
$1993-1997 $ (lag 5).
  }\label{fig:overlap1}
\end{figure}

\subsection{Temporal stability of the correlation
matrices}\label{tempstab}

Temporal stabilities of the matrices were investigated for annual
variation. In \cite{pre65a}, it was found that the largest four
eigenvectors obtained from high-frequency data (returns at 30-min
intervals) were stable up to time-lags of 1 year, while the
largest two eigenvectors from 30 years of daily data were stable
for time-scales up to 20 years.

In this investigation, we computed  overlap matrices with entries
given by estimated correlations between the leading eigenvectors
from the last $5-$yr epoch, $1998-2002$, with their analogues from
the preceding epochs, $1997-2001$ to $1993-1997,$ as follows: for
each epoch, the eigenvectors corresponding to the 15 largest
eigenvalues were chosen; each eigenvector was expanded to include
components for every share present in the epochs being compared
and when a share was not included in one of the epochs, a value of
zero was assigned \footnote{This takes into consideration the fact
that each epoch is comprised of slightly different collections of
shares}. We let $U(E) $ denote the $N \times 15$ matrices whose
columns are the leading 15 eigenvectors from the $E^{\it th}$
epoch, $E=6$ for $1998-2002$,  $E=5$ for $1997-2001$, etc. The
overlap matrices are hence given by
 $O(t,\tau)={U(t)}^T U(t-\tau)$, where $t$ denotes an epoch and
 $\tau$ denotes a lag in years.

Figure \ref{fig:overlap1} gives an indication of the temporal
stability of the eigenvectors associated with the largest
eigenvalues for Cases 1 and 3.

In Case 1 the correlation of the eigenvectors corresponding to the
largest eigenvalue is 1 for all lags; the correlation of the
eigenvectors corresponding to the second eigenvalues is negative
for lags 1 to 4 and the correlation of the eigenvectors
corresponding to the third eigenvalues is positive for lags 1 to
4.

In Case 3, temporal correlations of the eigenvectors corresponding
to the largest eigenvalue is alternately positive then negative
for successive lags; correlations for the second eigenvector are
positive except for time lags of 2 and 5 years where there are no
correlations; correlations for the third eigenvalue are
insignificant at delays of 1 and 2 years, then tend to be positive
for delays of 3-5 years; correlations between eigenvectors
corresponding to the next seven eigenvalues tend to be positive
for lags 1 to 5 years. The negative correlations in Case 3 are
consistent with the negative correlation between the financial and
resources sectors and the switch in performance of corresponding
indices following the market stresses in emerging markets in the
2$^{\it nd}$ half of 1998 (cf. Figure ~\ref{fig:hist}). This case
offers the greatest evidence of temporal stability. The overlap
matrices are all computed against Epoch number 6, 1998-2002, where
the market is also influenced by the crash of the Rand in 2001
(cf. ZAR/USD exchange rate inset in Figure ~\ref{fig:hist}).

There is also evidence of stability for Case 2 - in this case the
results are similar to but slighter weaker than those for Case 3.

\begin{figure}
  \centering
  \includegraphics[width=8.5cm]{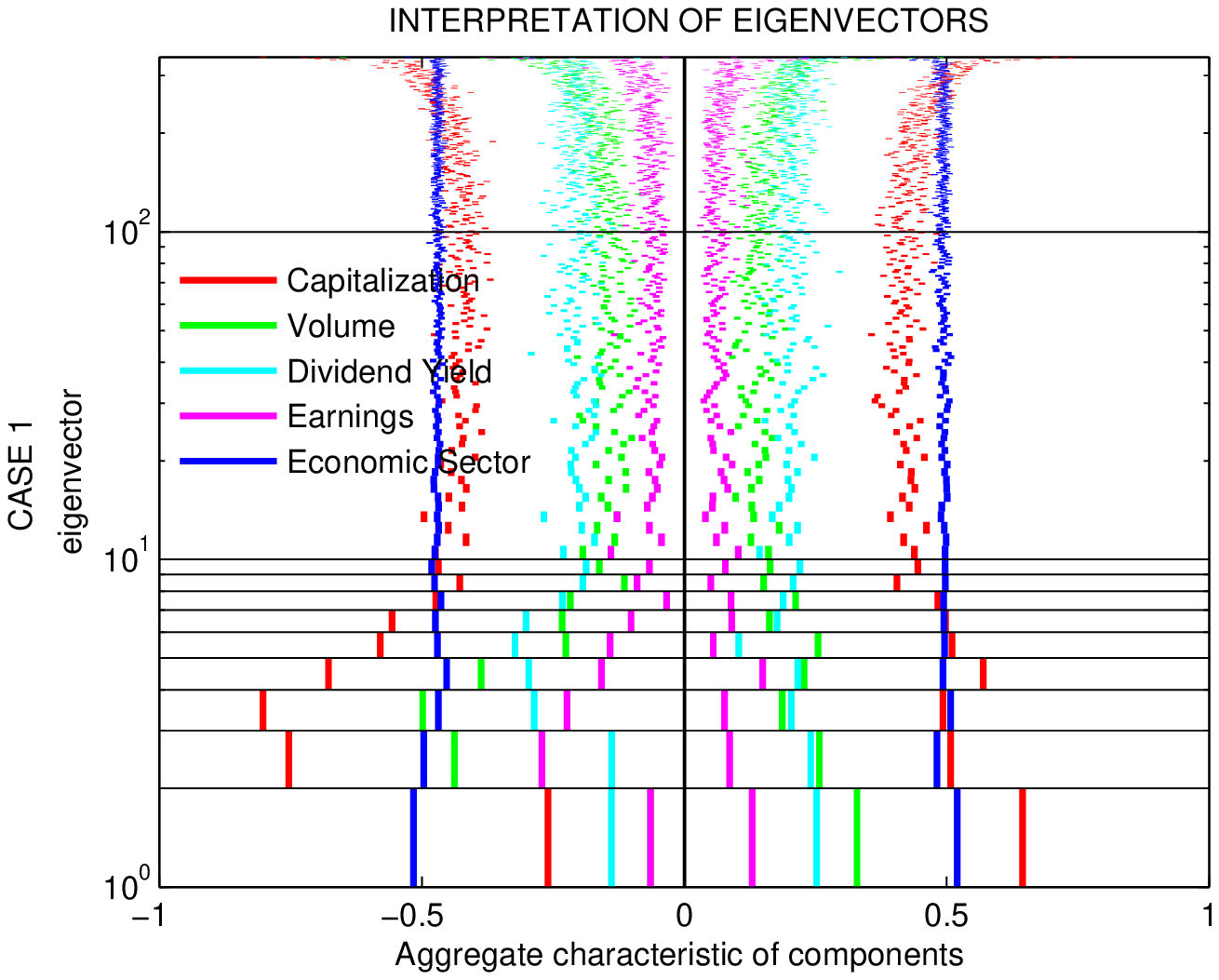}
  \includegraphics[width=8.5cm]{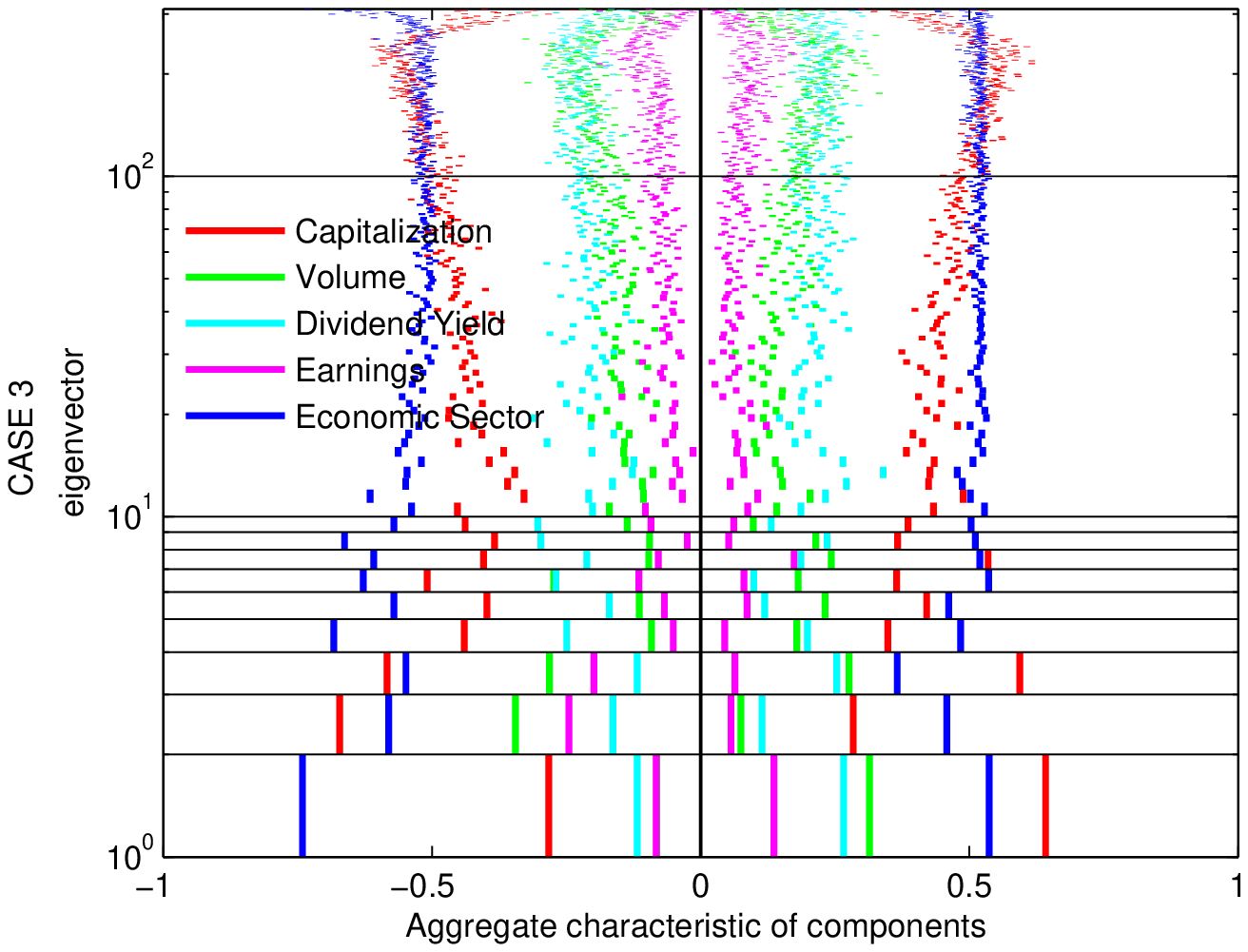}
  \caption{Fundamental characterizations of components of the
  eigenvectors for Cases 1 and 3 for 1998-2002 are given in Figure ~\ref{fig:char1} (a) and (b), respectively.
  Characteristic spectra are plotted in increasing order: the lowest band
  represents the spectrum for the first eigenvector, the next
  spectrum corresponds to the second eigenvector, etc.
  Characteristics used are: market capitalization, volume traded, dividend yield, earnings per share.
  For each fundamental characteristic, values were mapped to [0,1];
  negative values for components (on the left) indicate  short positions. }\label{fig:char1}
\end{figure}

\subsection{Interpretation of leading eigenvectors}

Several studies have investigated market segmentation or
clustering via metrics obtained from correlation matrices
(\cite{euphysj11a}, \cite{pre63a}, \cite{pre64a} and
\cite{BoCaLiMiVaMa}). In \cite{pre65a}, the authors are able to
interpret eigenvalues deviating from the RMT noiseband in a
similar way. In that investigation, because it was an order of
magnitude greater than the rest, the effect of the leading
eigenvalue was removed from the data by regressing each stock
return times series against the leading eigenmode to obtain a new
correlation matrix from stock specific return components (as in
the Capital Asset Pricing Model). The new leading eigenvector
exhibited significant contributions from about 1/3 of the 999
stocks, all with large values for market capitalization. The next
9 eigenvectors all contained stocks belonging to distinct economic
sectors.

In this investigation, the leading eigenvalue for Case 3 was found
to contribute much less significantly to the overall trace
compared to the rest of the large eigenvalues. Moreover, its
eigenvector components were not temporally stable, but displayed
anti-correlations over time as the behaviour of market
participants changed through the 1997 Russian GKO default, the
1998 emerging market contagion and the 2001 crash of the SA
currency.

Since the leading eigenvectors could not be identified with
distinct economic sectors, fundamental characteristics of each
eigenvector  were considered\footnote{{\em Fundamental} is used in
the 'bottom up' sense of financial analysis, i.e. in terms of
unique characteristics of stocks such as earnings, dividends,
market capitalization, book value, etc... }. Eigenvector
components were weighted according fundamental properties and
particular attention was paid to eigenvectors corresponding to the
largest eigenvalues. The fundamental properties considered were:
market capitalization, volume traded, dividend yield and earnings
 per share. These variables were normalized and mapped to numbers
between zero and unity. Economic sectors were interpreted in terms
of spectra, ranging from resources, through industrial,
non-cyclical and then cyclical shares into financial and then
technology shares, as classified by the JSE, from smaller values
to larger values and rescaled for the fundamental characteristic
graphs.

This representation offers a method to inspect the differences
between the compositions of the eigenmodes corresponding to
different eigenvectors.

Figure \ref{fig:char1} gives representations  of the fundamental
characteristics for eigenvectors corresponding to eigenvalues
ranging from largest (bottom) to smallest (top of range). We
include results for Cases 1 and 3.

This analysis, together with the tests of temporal stability,
suggests that the eigenmodes\footnote{The eigenmode for each
eigenvector is the timeseries derived from the timeseries of the
eigenvector components. Leading eigenmodes are also sometimes
referred to as {\it principal components}. } are not easily
interpreted in terms of isolated characteristics. Instead
eigenmodes may be viewed as being representative of distinct
trading strategies prevalent in the market itself. This conclusion
is motivated by the observation that negative component values of
the eigenvectors imply shorting and the positive values imply long
positions.

It can be deduced from Figure  \ref{fig:char1} that those
eigenvectors which are associated with eigenvalues lying in the
noise band seem to correspond to trading strategies which hold
roughly equal long and short positions in mid-capitalization
stocks. Case 1 and Case 3 are qualitatively the same in this sense
for eigenvalues in the noise band.

\begin{figure}
  \centering
  {\rotatebox{-90}{\includegraphics[width=8.5cm]{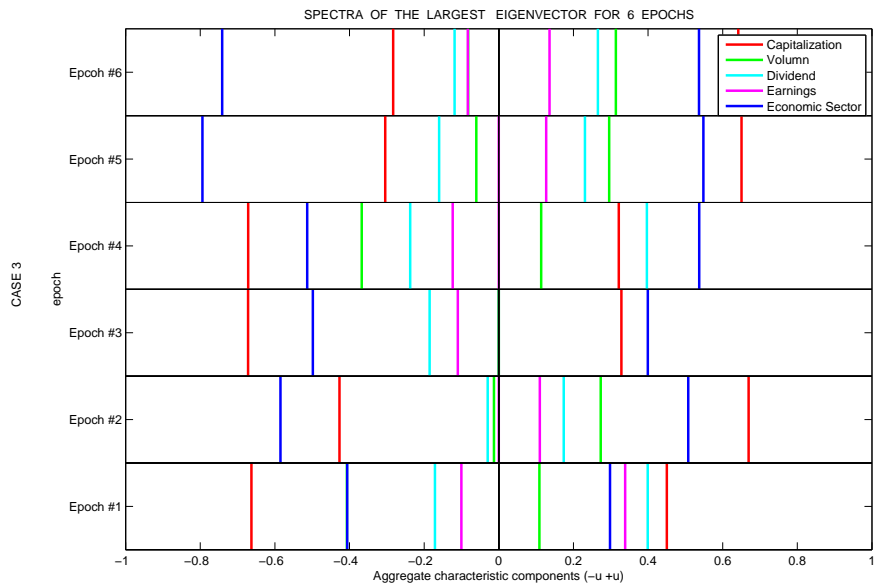}}}
  \caption{Fundamental characteristics of components of the $1^{st}$ eigenvectors for the 6 epochs for Case
  3, where Epochs 1 to 6 demarcate time windows 1993-1997 to 1998-2002 (cf. Figure ~\ref{fig:hist}). }\label{fig:char2}
\end{figure}

For Case 1, the first eigenvector is long in large capitalization,
large volume resource and industrial shares. The next leading
eigenvector carries short positions in large capitalization, large
volume shares. For all the eigenvectors in this case there is
indistinct economic sector characterisation.  This gives further
evidence that there is information loss with zero padding and
zero-order hold.

For Case 3 of Epoch 6, the leading eigenvector has similar long
positions as in Case 1. The second eigenvector is quite different
from  the first for Case 3, as well as from its counterpart in
Case 1: t exhibits long positions in stocks with smaller
capitalization and short positions in stocks with relatively
larger capitalization;  it is long in stocks with  low earnings,
low volume and low dividend yield. In general the leading
eigenvectors in Case 3 have more varied compositions.

Small eigenvalues for Case 3 correspond to trading strategies
which replicate long and short positions in small capitalization,
low earnings and lower volume and dividend yield relative to the
noise band. Cases 1 and 3 deviate markedly in this regard.

Similarly, inspection of the graph shows that characteristics for
the rest of the eigenvectors are different in Cases 1 and 3. In
the former, the characteristics seems to settle to a noisy
composition sooner. This is consistent with findings for the
inverse participation ratios.

In Figure \ref{fig:char2} we compare the fundamental
characteristics for the leading eigenmode for the different
epochs. In Epoch 1, the $1^{st}$ eigenmode is long in
comparatively higher earnings, smaller capitalization  and low
volume stocks, while it is short lower earnings and large
capitalization stocks. In contrast, in Epoch 5 (1997-2001) and
Epoch 6 (1998-2002), the $1^{st}$ eigenmodes are dominated by long
positions in large capitalization stocks and short positions in
small capitalization financial stocks, where the financial sector
corresponds to a economic sector value of 0.8. This depiction of a
shift in the market is consistent with the historical context
observed in Figure ~\ref{fig:hist}.

In Figure \ref{fig:char2} the short positions in Epoch 5 and 6
offer the only occurrences where aggregate quantification of
sector participation correlates uniquely with one sector. In
general, there is always significant share concentration in the
resources sector (sector value 0.0) for the period of
investigation as well as varying activity in industrial (sector
value 0.6) and financial stocks. As a result, eigenmodes are not
differentiated by sector participation and in particular, the
single aggregate quantifier for sector participation of eigenmodes
considered in this investigation is generally ineffective. This is
not surprising in an emerging market, where stock concentration,
currency volatility and generally high co-movement with commodity
prices blur economic sector independence. Inspection of these
quantifiers for the second eigenmodes corroborates these findings.

\section{Conclusions}

Our investigation exposes some notable differences in the spectral
properties of the correlation matrices computed by the three
different methods outlined. As in preceding analyses of financial
market data, in all cases we have found that the distribution of
eigenvalues  exhibits: (1) a significant part of the spectrum
falls within the range of random matrix predictions, and (2) there
exists  a small no. of large leading eigenvalues. However, we
found that by computing measured-data correlation matrices, Case
3, a far less substantial part of the spectrum falls within the
Wishart  range (Eqn. \ref{wishart}) than when computing
correlations with zero padding, Case 1. Similar results were found
when comparing the inverse participation ratios of Cases 1 and 3
with their RMT counterparts. Results for Case 2, which
incorporated zero-order hold but not zero padding, varied with the
RMT tests; here the eigenvector components and inverse
participation ratios were closest to RMT predictions.

Our investigation suggests that zero padding and zero-order hold
increases the level of noise in the estimation  of correlation
matrices. The correlations between leading eigenvectors from
successive epochs also showed evidence of greater temporal
stability when measured-data correlations were used.

Our fundamental characteristic investigation suggests that the
leading eigenmodes may be interpreted in terms of independent
trading strategies with long range correlations. These are more
distinct for the measured-data correlation case than when there
was zero padding and zero-order hold.

\section{Acknowledgements}
We thank the referees of Physica A for their pertinent comments
and questions. DW thanks National Research Foundation Thuthuka
Grant TTK2005081000005 and University of Cape Town Research
Council for financial support.


\end{document}